\documentclass[twocolumn,showpacs,preprintnumbers,amsmath,amssymb]{revtex4}

\usepackage{graphicx}
\usepackage{dcolumn}
\usepackage{bm}

\begin{document}

\title{Accessing the transport properties of graphene and its multi-layers 
\\at high carrier density}

\author{J. T. Ye$^{1,2*}$, M. F. Craciun$^{3}$, M. Koshino$^4$, S. Russo$^3$, S. Inoue$^2$, H. T. Yuan$^{1,2}$, H. Shimotani$^{1,2}$, A. F. Morpurgo$^5$, Y. Iwasa$^{1,2,6}$}
\email{yejianting@ap.t.u-tokyo.ac.jp, iwasa@ap.t.u-tokyo.ac.jp}

\affiliation{$^1$ Quantum-Phase Electronics Center and Department of Applied Physics, The University of Tokyo, 7-3-1 Hongo, Bunkyo-ku, Tokyo 113-8656, Japan}
\affiliation{$^2$ Institute for Materials Research, Tohoku University,
Sendai 980-8577, Japan}
\affiliation{$^3$ Centre for Graphene Science, University of Exeter, EX4 4QF Exeter, United Kingdom}
\affiliation{$^4$ Department of Physics, Tokyo Institute of Technology, 2-12-1 Ookayama, Meguro-ku, Tokyo 152-8551, Japan}
\affiliation{$^5$ DPMC and GAP, UniversitŽ de GenŽve, 24 quai Ernest Ansermet, CH1211 Geneva, Switzerland}
\affiliation{$^6$ CERG, RIKEN, Hirosawa 2-1, Wako 351-0198, Japan}

\date{\today}

\begin{abstract}
We present a comparative study of high carrier density transport in mono-, bi-, and trilayer graphene using electric-double-layer transistors to continuously tune the carrier density up to values exceeding 10$^{14}$ cm$^{-2}$. Whereas in monolayer the conductivity saturates, in bi- and trilayer filling of the higher energy bands is observed to cause a non-monotonic behavior of the conductivity, and a large increase in the quantum capacitance. These systematic trends not only show how the intrinsic high-density transport properties of graphene can be accessed by field-effect, but also demonstrate the robustness of ion-gated graphene, which is crucial for possible future applications.
\end{abstract}

\pacs{Valid PACS appear here}

\maketitle
Transport through graphene is currently investigated in the low carrier density regime ($\textit{n}\sim10^{12}$ $\textnormal{cm}^{-2}$), where electrons behave as unusual chiral particles \cite{1,2}. Despite exciting theoretical predictions (superconductivity \cite{5,6,7}) and clear technological relevance (transparent electrodes \cite{8}, supercapacitors \cite{9}, and bio-sensors \cite{10}), the high carrier density regime has remained vastly unexplored due to the limited density range accessible in conventional transistors \cite{1,2}. The recent development of so-called ionic-liquid gates, in which the coupling between gate electrode and transistor channel is realized through moving ions that form an electric double layer (EDL) at the liquid/channel interface (Fig. 1a), is now changing the situation. With ionic-liquid gates, the gate voltage applied Ðup to several voltsÐ drops across the approximately 1-nm thick EDL, leading to a very large geometrical capacitance. As a result, the induced carrier density can easily exceed $\textit{n}_{\textnormal{2D}}\approx10^{14}$ $cm^{-2}$, much larger than what is achieved in conventional solid-state field-effect transistors (FETs). Such a very strong field effect is valuable for technological applications (for instance, in organic FETs \cite{11}, where it enables low-voltage operation), and as a versatile and effective tool to tune electronic states in a rich variety of systems (by modulating metal insulator transition \cite{12}, magnetoresistance \cite{13}, and by inducing superconductivity at the surface of insulators \cite{14,15}).

\begin{figure}
\includegraphics[width=8.5 cm]{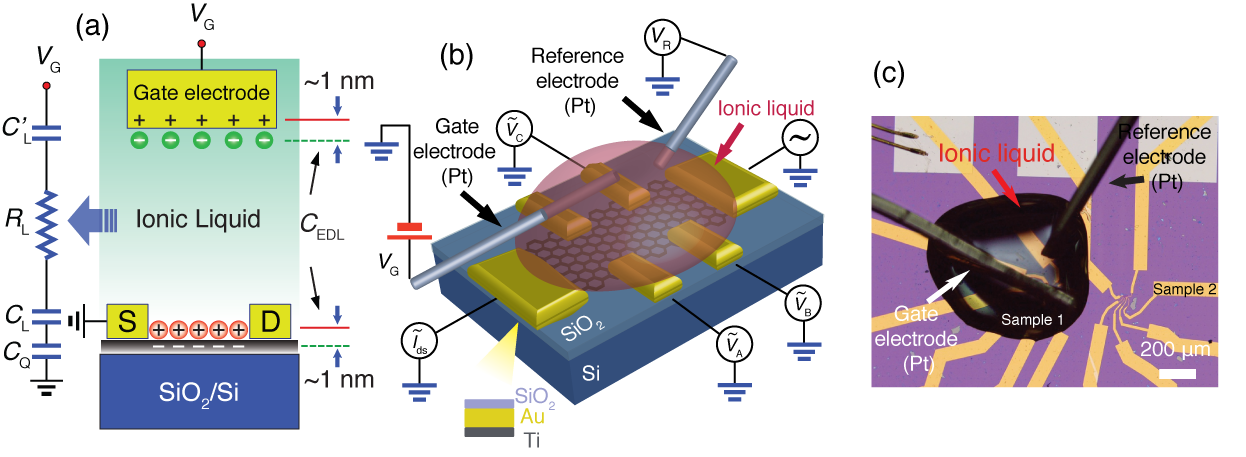}
\centering
\caption{\label{fig:epsart} (a) A schematic cross-section of graphene EDLT, with the equivalent electrical circuit (b) A schematic representation of a device including the bias configuration used in the electrical measurements. (c) Optical microscope image of an actual device. One of two graphene device is immersed in the ionic liquid together with two Pt wires, acting as gate and quasi-reference electrodes.
 }
\end{figure}

Recent work shows that ion gating can be used in combination with graphene. Experiments (\textit{e.g.}, Raman spectroscopy \cite{16}, quantum capacitance \cite{17}, \textit{etc.}) have focused almost exclusively on specific properties of monolayer at carrier density values up to  $\textit{n}_{\textnormal{2D}} \approx  5\times10^{13}$ $\textnormal{cm}^{-2}$, but no characteristic high carrier density features in the transport properties were identified. Here, as an efficient strategy to reveal these characteristic features, we perform a comparative study of transport in ion-gated mono-, bi-, and trilayer graphene at high carrier density of ~10$^{14}$
$\textnormal{cm}^{-2}$. The motivation for this strategy is twofold. First, when  $\textit{n}_{\textnormal{2D}}$ exceeds values of $~10^{13}$ $\textnormal{cm}^{-2}$, differences between monolayer and bi-/trilayer are expected, because in the latter systems the higher-energy split off bands start to be populated, which can provide an effective way to identify signatures of the intrinsic properties characteristic of the layers of different thickness. Second, in bi- and trilayer, the regime in which higher energy bands are populated has not yet been studied, and it is currently unknown how the opening of parallel transport channels affects the conductivity of these systems.

Mono-, bi-, and trilayer graphene devices were fabricated on SiO$_2$/Si substrates by exfoliating graphite (see Ref. \cite{Supp}) \cite{1}. A small droplet of ionic liquid was applied onto the devices, covering the graphene layer, the gate, and a quasi-reference electrode, as shown in Fig. 1(b) and (c). The droplet can be easily removed and substituted with a different ionic liquid, enabling the comparison of EDLTs realized on the same graphene layer, with different ionic liquids (Fig. S2a in Ref. \cite{Supp}): this is important to check that the features observed in the experiments are not artifacts caused by the specific ionic liquid chosen. For all devices, the longitudinal sheet resistivity $\rho_{\textnormal{xx}}$, and Hall coefficient $\textit{R}_{\textnormal{H}}$ were simultaneously measured at room temperature, in a Hall bar configuration, as a function of  $\textit{V}_{\textnormal{G}}-\textit{V}_{\textnormal{ref}}$ ($\textit{V}_{\textnormal{G}}$ is the voltage applied on the Pt gate electrode and $\textit{V}_{\textnormal{ref}}$ is the voltage measured on the quasi-reference electrode as shown in Fig. 1(b). We found $\textit{V}_{\textnormal{ref}}\approx0$ over the whole sweep range of $\textit{V}_{\textnormal{G}}$ (Fig. S2b in Ref. \cite{Supp}), which ensured almost all the applied$\textit{V}_{\textnormal{G}}$ dropped at the liquid/graphene interface). The measurements were performed in a limited $\textit{V}_{\textnormal{G}}$ range, to avoid the occurrence of chemical reactions between the ionic liquid and graphene, as it is necessary to obtain reproducible and reversible results (Fig. S3, S4 in Ref. \cite{Supp}). Despite this limitation, carrier density $\approx2\times10^{14}$ $\textnormal{cm}^{-2}$ could be reached.

Figure 2 shows the $\textit{V}_{\textnormal{G}}$ dependence of the sheet conductivity $\sigma_{\textnormal{2D}}=1/{\rho}_{\textnormal{xx}}$ for mono-, bi-, and trilayer graphene devices fabricated using ABIM-TFSI as ionic gate. For monolayer graphene (Fig. 2(a)), a linear increase of $\sigma_{\textnormal{2D}}$ is observed upon accumulating either electrons or holes, within $\Delta\textit{V}_{\textnormal{G}}\approx\pm1 V$ from the charge neutrality point (CNP). For larger $\textit{V}_{\textnormal{G}}$, $\sigma_{\textnormal{2D}}$ exhibits a pronounced saturation \cite{18}. The onset of a trend towards $\sigma_{\textnormal{2D}}$ saturation is normally seen in conventional SiO$_2$-based monolayer graphene FETs with sufficient high mobility \cite{19}. Here, the use of EDLT makes the phenomenon unambiguously clear, owing to the much larger carrier density range spanned. The behavior of bi- and trilayer graphenes (see Fig. 2(b) and (c)) differs from that of monolayer. In particular, the linear increase of $\sigma_{\textnormal{2D}}$ appears within a narrower $\textit{V}_{\textnormal{G}}$ range of $\Delta\textit{V}_{\textnormal{G}}\approx\pm0.5 V$  near the CNP. More distinctly, outside this range the $\sigma_{\textnormal{2D}}$ exhibits a non-monotonic behavior Ðboth for electrons and holesÐ before keeping increasing further at higher $\textit{V}_{\textnormal{G}}$.

Since charge accumulation in graphene EDLT devices is not simply described by a geometrical capacitance (see below), it is necessary to independently determine the sheet carrier density $\textit{n}_{\textnormal{2D}}$ as a function of $\textit{V}_{\textnormal{G}}$, in order to interpret the $\sigma_{\textnormal{2D}}$ data. To this end we have simultaneously measured the Hall resistance as a function of $\textit{V}_{\textnormal{G}}$. The blue lines in Fig. 2(d)-(f) display the resulting $\textit{n}_{\textnormal{2D}}$ derived for devices fabricated on layers of different thickness. As expected, at the value of $\textit{V}_{\textnormal{G}}$ corresponding to the $\sigma_{\textnormal{2D}}$ minimum $\textit{n}_{\textnormal{2D}}$ changes sign, confirming the shift of Fermi level $\textit{E}_{\textnormal{F}}$ across the CNP (Fig. S3 in Ref. \cite{Supp}). Using $\textit{n}_{\textnormal{2D}}$ determined from Hall measurements, we can directly extract the mobility of the devices. We find maximum values of 5.5, 3.5, and $9\times10^{3} \textnormal{cm}^{2}\textnormal{/Vs}$ close to the neutrality point, for mono-, bi-, and trilayer graphene respectively; similar mobility values are also observed using other ionic liquid, for instance DEME-TFSI (Fig. S5 in Ref. \cite{Supp}).

\begin{figure}
\includegraphics [width=8.5 cm] {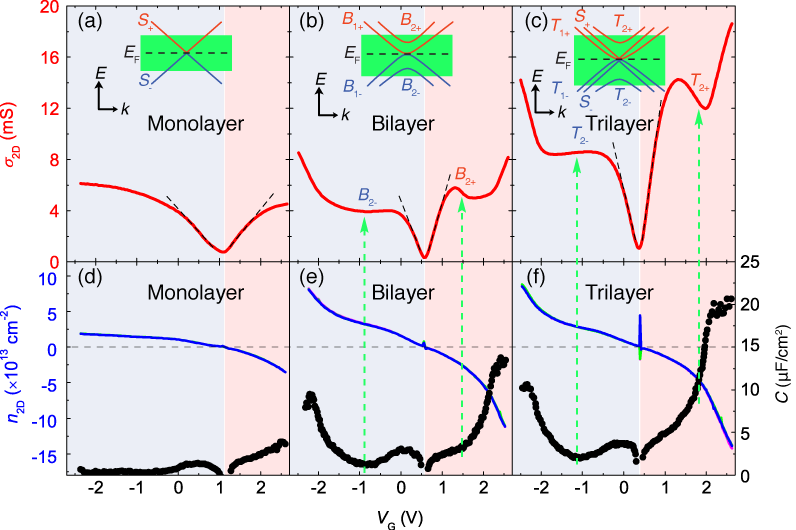}
\centering
\caption{\label{fig:epsart} (a)-(c) $\sigma_{\textnormal{2D}}(\textit{V}_{\textnormal{G}})$ of mono-, bi-, and trilayer graphene devices realized using ABIM-TFSI ionic-liquid gates. The black dashed lines indicate the linearity of $\sigma_{\textnormal{2D}}(\textit{V}_{\textnormal{G}})$ near the CNP (vertical white lines). (d)-(f) $\textit{n}_{\textnormal{2D}}$ as extracted from Hall measurements for graphene layers of different thickness, showing three perfectly overlapped data-sets measured at different magnetic field values (Fig. S3b in Ref. \cite{Supp}). Plot (d)-(f) also show the capacitance $\textit{C}$ of the layers. In panel (e) and (f), the vertical arrows show coincidence in $\textit{V}_{\textnormal{G}}$ of the anomalies in $\sigma_{\textnormal{2D}}$ and the increase in $\textit{C}$. The insets of panel (a), (b), and (c) illustrate  the band structure of mono-, bi-, and trilayer graphene. The green shaded areas illustrate the range within which the $\textit{E}_{\textnormal{F}}$ can be shifted.}
\end{figure}

\begin{figure}
\includegraphics [width=8.5 cm] {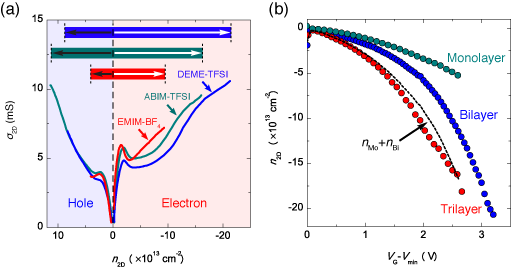}
\centering
\caption{\label{fig:epsart} (a)  $\sigma_{\textnormal{2D}}$ of bilayer graphene devices fabricated using three different ionic liquids as a function of $\textit{n}_{\textnormal{2D}}$ (measured from the Hall effect). The inset illustrates the maximum value of $\textit{n}_{\textnormal{2D}}$ for electrons (white arrows) and for holes (black arrows) accessible with the different ionic liquids. The largest $\textit{n}_{\textnormal{2D}}$ reached in this study is $2\times10^{14} \textnormal{cm}^{-2}$ using DEME-TFSI. (b) Dependence of $\textit{n}_{\textnormal{2D}}$ on $\textit{V}_{\textnormal{G}}$ (measured from the CNP) for mono-, bi-, and trilayer graphene devices, using DEME-TSFI as ionic liquid. The dashed black line is the sum of $\textit{n}_{\textnormal{2D}}$ of mono- and bilayer graphene, which compares well to the $\textit{n}_{\textnormal{2D}}$ measured in the trilayer. }
\end{figure}

Having determined the $\textit{n}_{\textnormal{2D}}$ of carriers, we extract the total capacitance of the devices from $\textit{C}=d\textit{n}_{\textnormal{2D}}/d\textit{V}_{\textnormal{G}}$. As shown in Fig. 2(d)-(f), when plotted as a function of $\textit{V}_{\textnormal{G}}$, a strong asymmetry between electrons and holes is clearly present, in sharp contrast to carrier accumulation using solid dielectrics. As electron-hole symmetry is known to approximately hold in graphene on the studied energy scale, we attribute the asymmetries observed to the properties of the EDLs. In large part, they originate from the different size of the positive and negative ions forming the ionic liquid, which are responsible for different thickness of the EDL for opposite polarities of the $\textit{V}_{\textnormal{G}}$. Indeed, the details of the asymmetry are different for different ionic liquids, as can be seen in Fig. 3(a) where the characteristics of a bilayer graphene device realized using ABIM-TFSI, DEME-TFSI, and EMIM-TFSI are shown (note also that when plotted as a function of $\textit{n}_{\textnormal{2D}}$, the asymmetry is considerably less pronounced). More importantly, however, the devices based on different liquids exhibit a fairly good agreement in the main features of the $\sigma_{\textnormal{2D}}$, including the absolute values, the non-monotonic behavior, and the position of the features as a function of $\textit{n}_{\textnormal{2D}}$. This indicates that the features in the $\sigma_{\textnormal{2D}}$ are intrinsic to bilayer, which are not influenced by the specific ionic liquid used. Similar insensitivity to the ionic liquids was also observed in mono- and trilayer.

It is apparent from Fig. 2 that $\textit{C}$ depends very strongly on $\textit{V}_{\textnormal{G}}$. This is because \textit{C}, the total capacitance  measured experimentally, is given by $1/\textit{C}=1/\textit{C}_{\textnormal{L}}+1/\textit{C}_{\textnormal{Q}}$, where $\textit{C}_{\textnormal{L}}$ is the ÒgeometricalÓ capacitance of the EDL formed between Pt and graphene, and $\textit{C}_{\textnormal{Q}}$ is the so-called quantum capacitance associated to the finite density of states (DOS) of graphene. Owing to the large $\textit{C}_{\textnormal{L}}$ of EDLs (several tens of ${\mu}\textnormal{F/cm}^{2}$) \cite{17}, $\textit{C}_{\textnormal{Q}}$ dominates the total $\textit{C}$, which is why the $\textit{E}_{\textnormal{F}}$ can be tuned by applying only small $\textit{V}_{\textnormal{G}}$ (in conventional graphene FETs, $\textit{C}_{\textnormal{Q}}$ is normally negligible, because the $\textit{C}_{\textnormal{L}}$ is three orders of magnitude smaller than of EDLs Ðtypically, for a 300 nm SiO$_2$ layer, ~12 nF/cm$^{-2}$). The dominant role of $\textit{C}_{\textnormal{Q}}$ naturally explains why the amount of carrier accumulated in graphene layers of different thickness (at the same $\textit{V}_{\textnormal{G}}$ and using a same ionic liquid) is very different, as a direct consequence of the different DOS.

The insets of Fig. 2(a)-(c) illustrate the main features of the band structures of graphene mono-, bi-, and trilayer. Whereas in monolayer only two (valence and conduction) linearly dispersing bands $\textit{S}_{2\pm}$ touching at zero energy are present, bi- and trilayer graphenes have additional bands at higher energy, around 0.4 ($\textit{B}_{2\pm}$) and 0.5$\sim$0.6 eV ($\textit{T}_{2\pm}$) from the CNP, respectively (Sect. 3 in Ref. \cite{Supp}); filling of these bands occurs when the $\textit{n}_{\textnormal{2D}}$ exceeds $\sim2$ and $7\times10^{13} $ $\textnormal{cm}^{-2}$ in these two cases (estimated from known band structure by neglecting changes induced by the perpendicular electric field generated by applied $\textit{V}_{\textnormal{G}}$). Indeed, the ÒanomaliesÓ (i.e., the non-monotonic behavior) in the $\sigma_{\textnormal{2D}}$ occur at density values, larger for trilayer than for bilayer, close to the ones estimated above. We therefore attribute the anomaly in $\sigma_{\textnormal{2D}}$ to the presence of a new scattering channel --inter-band scattering-- that opens when the higher bands are filled, and that, as it is known from conventional semiconductors \cite{20}, can strongly reduce the carrier mobility. This is also why no anomaly is seen in the monolayer, since in that case no higher energy band is present.

The interpretation of the $\sigma_{\textnormal{2D}}$ anomalies in terms of inter-band scattering is further supported by the observed behavior of the capacitance of bi- and trilayer graphene (Fig. 2(e) and (f)), which exhibits a sharp increase concomitant with the $\sigma_{\textnormal{2D}}$ anomalies since when the Fermi level enters the $\textit{B}_{2\pm}$ or $\textit{T}_{2\pm}$ bands the DOS increases in a step-like manner, and so does the $\textit{C}_{\textnormal{Q}}$ (in the experiments the step-like increase is broadened by disorder). Finally Fig. 3(b) shows the relation between $\textit{n}_{\textnormal{2D}}$ and \textit{V}$_{\textnormal{G}}$ for mono-, bi-, and trilayer devices, from which we find that $\textit{n}_{\textnormal{2D}}$ for the trilayer closely matches the sum of the densities in the mono- and bilayer. Such a relation is expected to approximately hold from the theoretical band structure of these systems, since the two lowest energy bands in trilayer correspond approximately to the linear band of monolayer and the lowest energy quadratic band of bilayer (Sect. 3 in Ref. \cite{Supp}).

\begin{figure}
\includegraphics [width=6.5 cm] {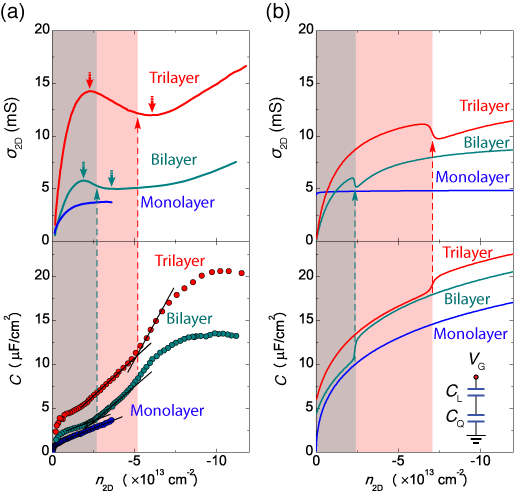}
\centering
\caption{\label{fig:epsart} (a) Measured data of $\sigma_{\textnormal{2D}}$ and $\textit{C}=d\textit{n}_{\textnormal{2D}}/d\textit{V}_{\textnormal{G}}$  as a function of $\textit{n}_{\textnormal{2D}}$ for mono-, bi-, and trilayer graphene. The red and cyan pair of arrows point to the $\sigma_{\textnormal{2D}}$ anomaly (non-monotonic dependence of $\sigma_{\textnormal{2D}}$ on $\textit{n}_{\textnormal{2D}}$). The dashed lines indicate how the $\sigma_{\textnormal{2D}}$ anomalies in bi- and trilayer, corresponds to the $\textit{n}_{\textnormal{2D}}$ where the capacitance increases. (b) Results of theoretical simulations for the same quantities plotted in panel (a). To analyze the data theoretically, we have chose for the geometrical capacitance a realistic value of $\textit{C}_{\textnormal{L}}= 40 {\mu}{\textnormal{F/cm}}^{2}$.}
\end{figure}

To substantiate all the considerations just made, we have analyzed the high-density electronic properties of mono-, bi-, and trilayer graphene in terms of a simple, well-defined theoretical mode \cite{18}. As our only goal at this stage is to show that all the main features observed in the $\sigma_{\textnormal{2D}}$ and $\textit{C}$ measurements naturally arise from a formally correct theoretical framework, for simplicity we have included only the effect of weak short-range scattering \cite{18} and adopted a self-consistent Born approximation scheme \cite{21} (for a precise quantitative analysis, other scattering mechanisms should be included as well, e.g., to capture the effect of long-range Coulomb potential \cite{23}, resonant scattering \cite{25}, as well as ripples \cite{26}). The experimental data for devices realized using ABIM-TFSI are compared with theory in Fig. 4. At a qualitative level, the theory reproduces the trends seen in both $\sigma_{\textnormal{2D}}$($\textit{n}_{\textnormal{2D}}$) and $\textit{C}$($\textit{n}_{\textnormal{2D}}$) curves, for the graphene layers of different thickness. This includes the trend towards $\sigma_{\textnormal{2D}}$ saturation in the monolayer, and the anomalous features in the $\sigma_{\textnormal{2D}}$ and $\textit{C}$ in bi- and trilayer. In particular, the non-monotonic behavior of the calculated $\sigma_{\textnormal{2D}}$ in bi- and trilayer can be traced back to the presence of interband scattering, which agrees with our initial interpretation and confirms the relevance of this process at high $\textit{n}_{\textnormal{2D}}$. Note, finally, that in the model we have not included the modification of the band structure due to the perpendicular electric field generated by $\textit{V}_{\textnormal{G}}$ (\textit{i.e.} the opening of a gap in bilayer \cite{27}, and the modification of the band overlap in trilayer \cite{28}); inclusion of these effects may lead to a better quantitative agreement for the values of $\textit{n}_{\textnormal{2D}}$ at which higher energy bands are populated in bi- and trilayer.

From the results obtained, we conclude that ionic-liquid gating is an effective and reliable technique to accumulate very large amounts of carriers in graphene-based materials. The $\textit{n}_{\textnormal{2D}}$ that have been achieved in this study are not far from those for which superconductivity is observed in graphite intercalated compounds (\textit{e.g.}, KC$_{8}$, $\sim 5\times10^{14}$ cm$^{-2}$) and, indeed, superconductivity in graphene at high $\textit{n}_{\textnormal{2D}}$ has been predicted theoretically \cite{5,6,7}. It is likely that these higher $\textit{n}_{\textnormal{2D}}$ values can be reached by biasing $\textit{V}_{\textnormal{G}}$ over a broader range at lower temperature \cite{15,29}. For applications, the high $\textit{n}_{\textnormal{2D}}$ leading to very high $\sigma_{\textnormal{2D}}$ values, and the good compatibility of ionic-liquid gates with graphene are important for the realization of transparent electrodes in flat panel displays \cite{8}, supercapacitors \cite{9} and biosensors \cite{10}. At the current stage, it appears that all the ingredients necessary for a rapid progress in directions of fundamental and technological interest that combine the unique properties of graphene and ionic-liquid gates are already available.

We thank D. Chiba, Y. Ohno, F. Matsukura, and H. Ohno for the electron beam facility and T. Ando for theoretical discussion. MFC, SR, and AFM would like to thank S. Tarucha for support. This research is funded by MEXT and JST. AFM gratefully acknowledges MaNEP and the Swiss National Science Foundation (project 200021\_121569) for financial support.

\bibliography{GrapheneEDLT.bib}
\end{document}